# Growth of Carbon Nanotubes on Quasicrystalline Alloys

Deep Jariwala[1*], Kaushik Chandra[1,2], Anyuan Cao[3], Saikat Talapatra[4] , Mutshihiro Shima[3], D. Anuhya[1], V.S.S.S. Prasad[1], R. Ribeiro[5], P.C.Canfield[5], D. Wu[5], Anchal Srivastava[6], R. K. Mandal[1], A. K. Pramanick[7], Robert Vajtai[2,3], Pulickel. M. Ajayan[2,3*], and G.V.S. Sastry[1,3♦]

[1] Department of Metallurgical Engineering, Banaras Hindu University, Varanasi-221005, India
[2] Mechanical Engineering and Materials Science department, Rice University, Houston, TX-77005.
[3] Department of Materials Science and Engineering, Rensselaer Polytechnic Institute, Troy NY, 12180, USA.
[4] Department of Physics, Southern Illinois University Carbondale, Carbondale, IL, 62901, USA.
[5] Ames Laboratory, Iowa State University, Ames, Iowa 50011, USA
[6] Department of Physics, Banaras Hindu University, Varanasi-221005, India.
[7] National Metallurgical Laboratory, Jamshedpur- 634, India

## Abstract

*We report on the synthesis of carbon nanotubes on quasicrystalline alloys. Aligned multiwalled carbon nanotubes (MWNTs) on the conducting faces of decagonal quasicrystals were synthesized using floating catalyst chemical vapor deposition. The alignment of the nanotubes was found perpendicular to the decagonal faces of the quasicrystals. A comparison between the growth and tube quality has also been made between tubes grown on various quasicrystalline and $SiO_2$ substrates. While a significant MWNT growth was observed on decagonal quasicrystalline substrate, there was no significant growth observed on icosahedral quasicrystalline substrate. Raman spectroscopy and high resolution transmission electron microscopy (HRTEM) results show high crystalline nature of the nanotubes. Presence of continuous iron filled core in the nanotubes grown on these substrates was also observed, which is typically not seen in MWNTs grown using similar process on silicon and/or silicon dioxide substrates. The study has important implications for understanding the growth mechanism of MWNTs on conducting substrates which have potential applications as heat sinks.*

## Introduction

Carbon nanotubes (CNTs) have captured the attention of materials scientists and technologists due to their unique one-dimensional structure by virtue of which they acquire superior electrical, mechanical and chemical properties. Numerous applications in diverse fields of technology are therefore envisioned for CNTs[1-4]. Their high current carrying capacity and amenability to doping enhances their use as interconnects in nanodevices and as semiconducting devices themselves. CNT growth on $SiO_2$ / patterned $SiO_2$ substrates and their surface selective growth mechanism has been studied extensively in recent experiments[5-7]. For many potential applications such as field emission displays, sensors and electronic devices it is important to grow CNTs on conducting substrates. The development of techniques for aligned growth of CNTs on suitable substrates has received special attention for their possible applications in displays, gigascale interconnects and micro/nanosensors[8-22]. However, well architectured growth of CNTs on metallic substrates that are highly conducting unlike $SiO_2$ has been a major challenge and is a highly desired goal of the field of research and technology development in CNTs. New experimental findings suggest that various metal alloys can be used as substrates for aligned nanotube growth[23-29]. However, the above reports pertaining to CNT growth on various metals/alloys have remained limited to those containing a good amount iron, cobalt or nickel which have been conventionally used as nanoparticle-catalysts for CNT growth. CNT growth on other conducting substrates is still elusive and yet to be properly understood and realized. In this article we report that quasicrystalline[30] alloys can be used as suitable substrate material for aligned nanotube growth. We have performed growth of aligned multiwalled carbon nanotubes (MWNTs) on faces of conducting decagonal quasicrystals (a dc conductivity of the order of 150-200 $(\Omega cm)^{-1}$ for a typical icosahedral quasicrystal, Al-Cu-Fe[31]) (DQCs)[32-33] alloy by floating catalyst chemical vapor deposition (CVD) method. A major advantage of choosing such a type of substrate, apart from it being electrically conducting, is the periodicity along ten-fold direction and quasiperiodicity in the plane perpendicular to it. This makes the decagonal quasicrystal anisotropic in many properties. For example, the electrical conductivity values differ by an order of magnitude in the two directions (A DQC Al-Pd-Mn has a resistivity of $1000\mu\Omega cm$ at 270K in the quasiperiodic direction[34]). This offers the possibility of preferred conducting paths when used as a device component. Another important aspect for the present choice is that the DQCs are reported to display enhanced catalytic properties[35-39].There are several binary[40] and multicomponent quasicrystals and the compositions of occurrence of a quasicrystalline phase can also be predicted using simple criteria[41].

∗ Present address: Department of Materials Science and Engineering, Northwestern University, Evanston, IL-60208, USA.
♦ Corresponding Author: G.V.S. Sastry: gvssastry.met@itbhu.ac.in , P. M. Ajayan: ajayan@rice.edu

The present study is aimed at exploring the possibility aligned growth of carbon nanotubes on different quasicrystalline substrates and relate their quality to the type of quasicrystal.

**Experimental Details**

Well-characterized DQC[42] and isocahedral quasicrystal (IQC)[43], substrates were selected for the substrate material. All the quasicrystalline investigated in this paper are listed in Table 1 along with their composition, type and their quasicrystalline phase stability temperatures. An optical micrograph of a representative tiny decaprism of DQC ($Al_{68.85}Co_{15}Cu_{6.15}Ni_{10}$) is shown in Fig. 1(a). While the external decaprismatic morphology may be a first indication of the occurrence of a quasiperiodic phase, the presence of a true 10-fold electron diffraction pattern from such phase (see Fig.1(b)) identifies a decagonal quasicrystal. This 10-fold symmetry axis is present along the periodic axis and normal to the quasiperiodic planes in the decagonal quasicrystal. This diffraction pattern bears some resemblance to the 5-fold pattern in the case of an icosahedral quasicrystal. Perpendicular to this 10-fold direction two sets of 2-fold axes are present along which the decagonal quasicrystal exhibits characteristic 2-fold diffraction patterns, Fig. 1(c) & (d), that are generally referred to as D and P patterns[44]. These are so named to represent the decagonal and pentagonal patterns, as the D pattern identifies the true characteristics of the decagonal quasicrystal (the particular periodicity present along the 10-fold axis and the quasiperiodic nature perpendicular to it) and the P pattern is similar to the 2-fold pattern of an icosahedral quasicrystal[44]. Decarods of 0.2-0.5 mm in section and 1-2 mm long were transferred to a horizontal CVD furnace to grow nanotubes. In addition melt spun ribbons of decagonal and icosahedral quasicrystals were also used for growth. We used the mixture of 0.4 g ferrocene in 40 ml xylene as the catalyst precursor[25] and carbon source at a reaction temperature of 775-800 ºC, and the CVD setup reported by Andrews[5] previously. A brief description of the procedure is as follows: The ferrocene/xylene solution (0.01 g/ml) was put into a source bottle connected to the inlet of a quartz tube housed in the CVD furnace, and preheated to 150 ~ 180 °C to keep the solution boiling inside the bottle. After the furnace was heated to 800 °C under a 200 mTorr pressure, we opened the source bottle valve to allow the solution vaporize into the quartz tube and grow nanotubes on the crystals and ribbons. An Si/300 nm $SiO_2$ substrate was kept alongside all quasicrystals to compare the growth. Typical reaction time was about 10 minutes. The structural characterization of the samples was performed using scanning electron microscopy (SEM) (Quanta 200FEG and ZEISS, EVO LS 25 ESEM), transmission electron microscopy (TEM) (Tecnai 20G$^2$ and JEOL JEM 200 CX) and Raman spectroscopy (Renishaw Raman microscope). The TEM studies were carried out on JEOL JEM 200 CX microscope at 120 kV and Tecnai 20G$^2$ operating at 200kV. The CNT samples were loaded on holey carbon grids and were covered with copper grids for protection. The deposits were scraped off the substrates and loaded onto TEM grids. Methanol was later dropped on to the covered grids to ensure proper dispersion of the sample.

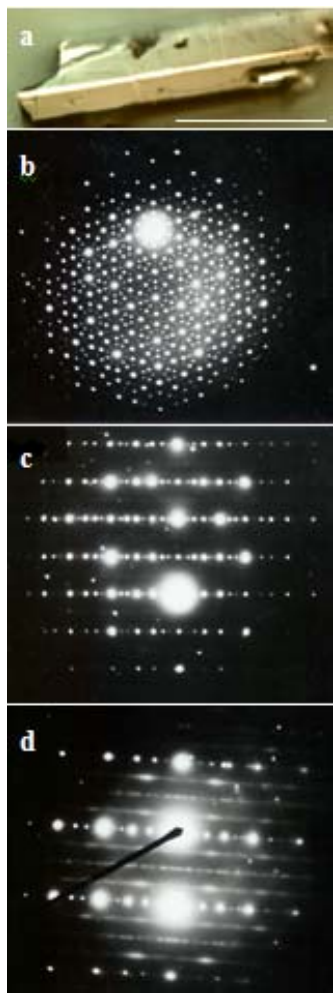

**Figure 1.** (a) An optical micrograph of a representative tiny decaprism is shown. (b) the presence of a true 10-fold electron diffraction pattern from a quasiperiodic phase is presented. (c) & (d) electron diffraction showing D and P patterns obtained from the quasicrystal.

**Results and Discussion**

Raman spectroscopy was carried out on the growth samples using a Renishaw Raman microscope. The excitation wavelength used for the spectroscopy was 514 nm which was generated using Ar ion laser. The least count of the instrument is 0.7 nm. Raman spectra of MWNTs on $SiO_2$ showed sharp G, D and 2D peaks of the graphitic lattice which are typical for MWNTs as reported in literature before[1, 45-48]. The G peak is observed at 1557.3 cm$^{-1}$.The D band peak associated with lattice disorder is observed at 1351.39 cm$^{-1}$ and consequently the 2D peak is observed at 2699.8 cm$^{-1}$. The $I_D/I_G$ ratio is approximately 0.522. This means that the MWNTs synthesized are of very high quality and the ratio is comparable to some of the values reported in literature[46-47]. Fig (2.a) shows the Raman spectrum of MWNTs on $SiO_2$ substrates. Fig. 2 (b), (c) and (d) shows the corresponding SEM images of MWNTs grown on $SiO_2$ substrates. A continuous mat of nanotubes is observed. Fig 3 shows results of our SEM analysis and Raman spectroscopy of the multi walled nanotubes on the DQC surface. In Fig. 3(a) a low magnification SEM image of a $Al_{68.85}Co_{15}Cu_{6.15}Ni_{10}$ decarod is shown. A continuous uniform mat of nanotubes covering the whole decarod is seen in this image. Raman spectrum, measured from this sample is presented in the inset. the D peak is seen around 1345 cm$^{-1}$ and the G peak is seen around 1588 cm$^{-1}$. The $I_D/I_G$ ratio is comparable to tubes grown on $SiO_2$ indicating no degradation of tube quality with change of substrate from insulating amorphous to conducting quasicrystalline. In Fig. 3 (b) and (c) higher magnification images of the nanotubes are shown. A high degree of alignment of the nanotubes was observed. It can be seen that the MWNTs grow on all the faces (perpendicular to them) of the DQC and cleaved away from each other at the edges of the decarods. The samples, with certain regions of the CNT growth were intentionally scraped off, and examined in SEM to explore the condition of the

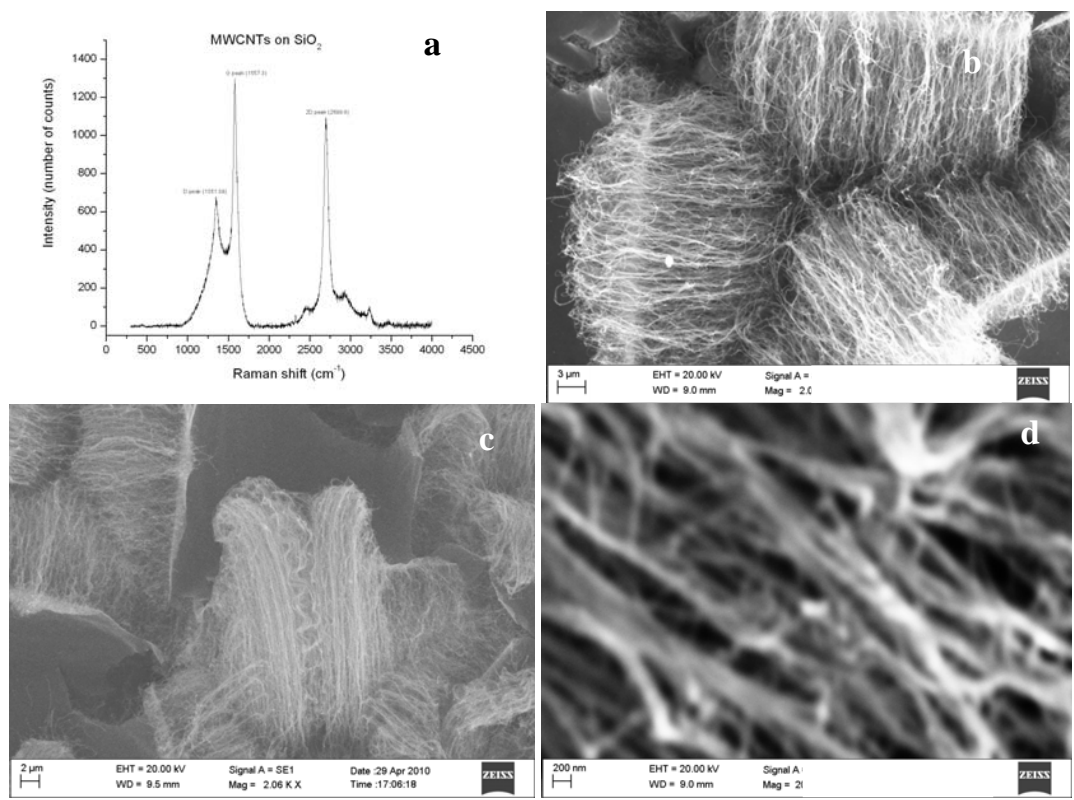

**Figure 2.** (a) Raman spectrum of MWNTs grown on Si/300nm SiO$_2$ substrates showing three distinct peaks namely the G (1557.3 cm$^{-1}$), D (1351.39 cm$^{-1}$) and 2D (2699.8 cm$^{-1}$) peaks which are characteristic of graphitic carbon nanotubes.(b) & (c) Low magnification SEM images of aligned MWNT bundles. (d) High magnification SEM image of the bundles showing individual tubes and alignment.

substrate surface. Figure 2d shows such an area (shown by the arrow) where the surface was found to be clean and devoid of any oxide. Similar growth was carried out on other DQC substrates of varying composition and also Icosahedral quasicrystalline (IQC) substrates. Growth was also carried out a decagonal single crystal of Al$_{73.2}$ Ni$_{10.7}$ Co$_{16.1}$ with 10 fold orientation which gave similar results. The substrate composition, type and their stability temperatures are provided in Table 1. The growth carried out DQCs with two other compositions namely Al$_{67}$ Co$_{15}$ Cu$_8$ Ni$_{10}$ and Al$_{67}$ Co$_{15}$ Cu$_8$ Ni$_{15}$ also yielded MWNT growth.. The SEM image of the nanotubes on these substrates are shown in Fig. 4(a). The SEM image shows tube like growth however, it doesn't appear to be a continuous mat of vertically aligned tubes. It rather appears as a sparse stunted tube growth. Similarly, ternary (Mg$_{32}$(Al, Zn)$_{49}$) and quaternary (Al$_{67}$Ga$_4$Pd$_{21}$Mn$_8$) IQC substrates were also tried to observe CNT growth. However, in both these cases, since the ternary IQC phase is unstable at the growth temperatures of about 800 ˚C the sample was placed in a colder region of the furnace where the temperature was just below the stability temperatures of the icosahedral phases. While this ensures that the icosahedral phase remains stable during growth, it reduces the chances of growth since the temperature of the substrate is lesser than typical growth temperatures which is indeed what we find. Fig. 4(b) shows SEM image of the deposit on IQC (Mg$_{32}$(Al, Zn)$_{49}$). The SEM image does not show any tube like structure and appears to be amorphous carbon. Similarly on the quaternary IQC (Al$_{67}$Ga$_4$Pd$_{21}$Mn$_8$) no distinct signs of aligned

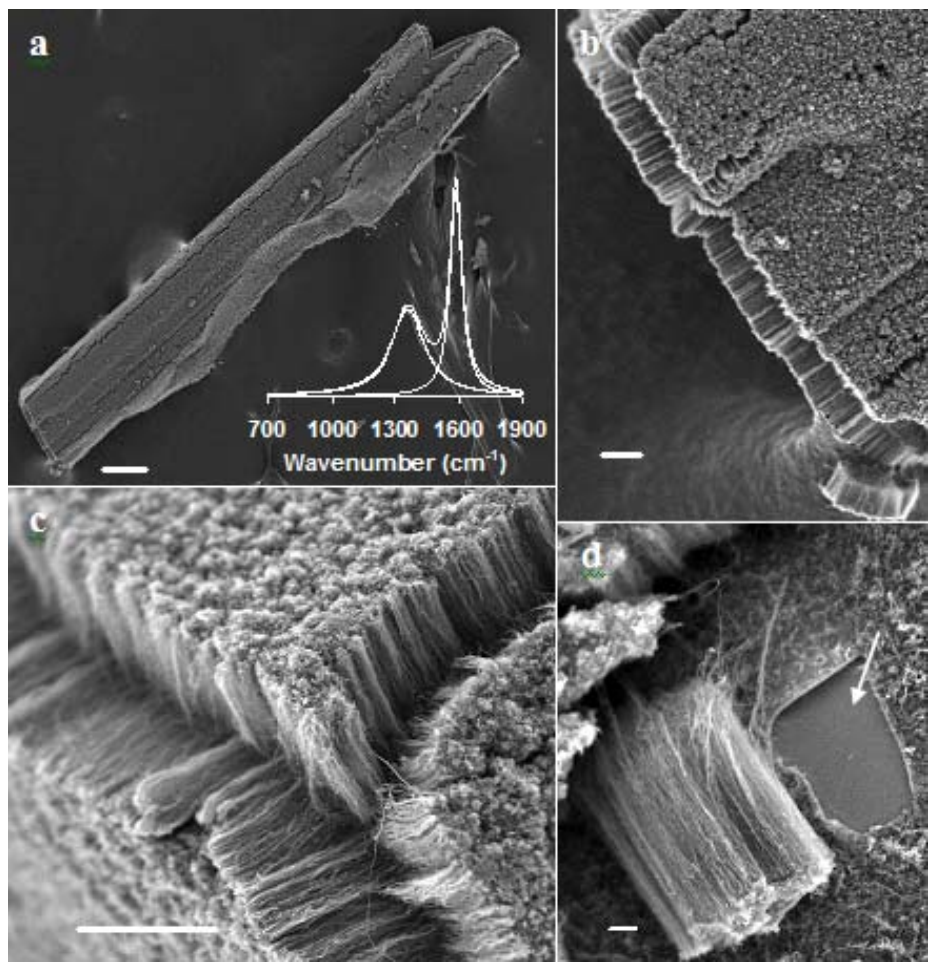

**Figure 3.** (a) Low magnification SEM image of a $Al_{68.85}Co_{15}Cu_{6.15}Ni_{10}$ DQC decarod showing continuous uniform aligned matt of nanotube growth on the faces of the quasicrystal (scale bar = 100 μm). (Inset) First-order Raman spectrum, measured from this sample showing peaks around 1345 cm$^{-1}$ (D) and a peak around 1588 cm$^{-1}$ (G). (b) & (c) higher magnification images of the nanotubes showing a high degree of alignment perpendicular to the faces of the DQC (scale bar = 10 μm). (d) Scraped off CNT growth area on the DQC surface showing the surface was clean and devoid of any oxide (scale = 1 μm).

MWNTs were visible. However, since the icosahedral phase stability was higher and closer to growth temperatures (~680 °C) some tubular structures were seen as shown in the SEM image in Fig. 4 (c) and (d). These tubular structures do not exactly look like nanotubes but more like amorphous carbon nanofibers. They also show nodular and/or helix-like morphology. Such structures have been reported recently and have potential applications in hydrogen storage[49].

**Table 1**

| S. No | Alloy composition | Type of quasicrystal | Stability range |
|---|---|---|---|
| 1. | $Al_{67}Co_{15}Cu_8Ni_{10}$ | Decagonal | < 900 °C[51-52] |
| 2. | $Al_{67}Co_{15}Cu_8Ni_{15}$ | Decagonal | < 900 °C[51-52] |
| 3. | $Al_{68.85}Co_{15}Cu_{6.15}Ni_{10}$ | Decagonal | < 900 °C[42] |
| 4. | $Al_{73.2}Ni_{10.7}Co_{16.1}$ | Decagonal | <900 °C[53] |
| 5. | $Mg_{32}(Al, Zn)_{49}$ | Icosahedral | < 500 °C[54] |
| 6. | $Al_{67}Ga_4Pd_{21}Mn_8$ | Icosahedral | <680 °C[43] |

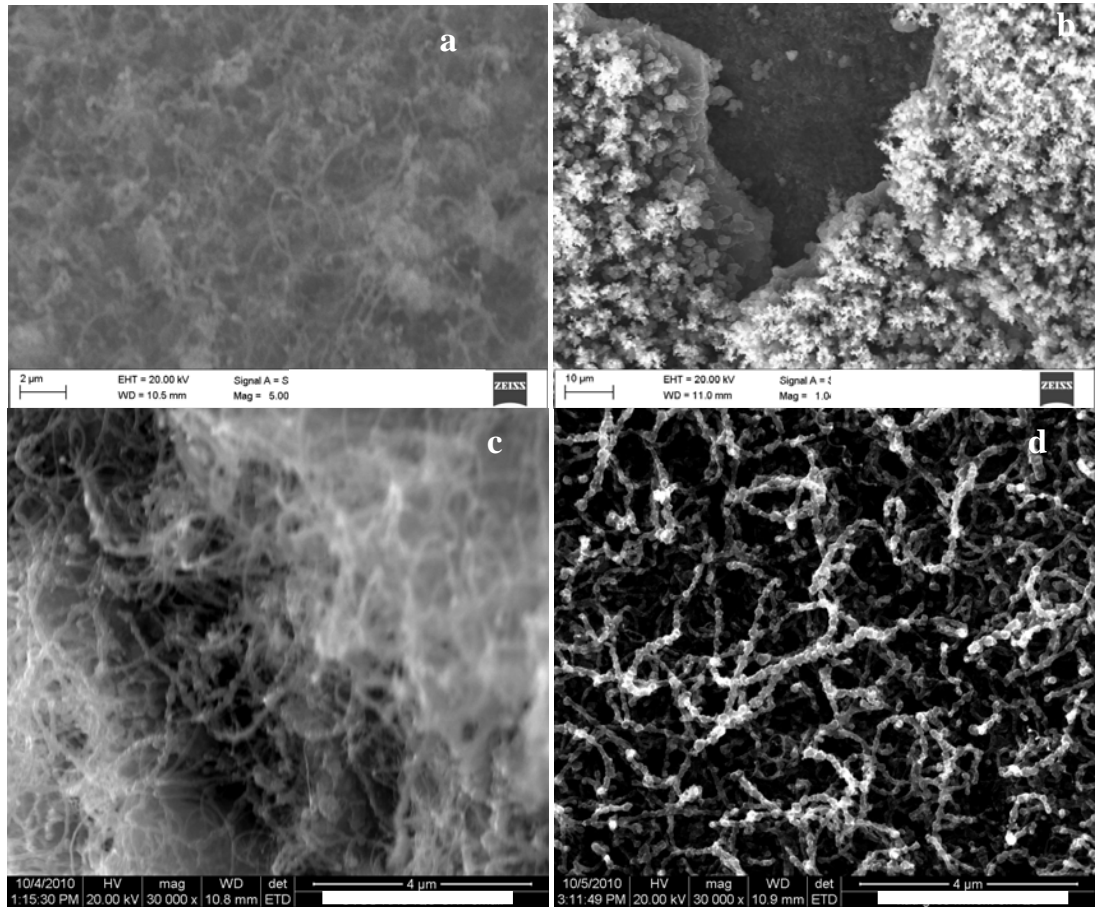

**Figure 4.** (a) SEM image of MWNT growth on $Al_{67}$ $Co_{15}$ $Cu_8$ $Ni_{10}$ DQC substrate showing stunted tubular growth.(b) SEM image of growth on ternary $Mg_{32}(Al, Zn)_{49}$ IQC substrate showing no tube like growth. (c) and (d) are SEM images of helical tube like growth on quaternary IQC $Al_{67}Ga_4Pd_{21}Mn_8$.

The multi-wall structure of the CNTs and its perfectness needs to be verified. However, a precise confirmation can only be obtained by high resolution transmission electron microscopy (HRTEM). Fig. 5(a) shows an HRTEM image of a single strand of a MWNT at low magnification grown on DQC decarod. Note the presence of a continuous filament in the core of the nanotube. The ferrocene-xylene CVD growth of CNTs often times results in encapsulated metal clusters which are γ-Fe[50]. Since the growth process uses iron catalyst and the amount of material needed to fill the core of the CNTs are substantial, we believe that the core is predominantly filled with Fe from the catalyst (however, presence of Co or Ni in solution cannot be ruled out only on the basis of electron diffraction patterns since they are present in the DQC substrate). The crystalline nature of this filamentary iron at the core can be inferred from the diffraction contrast (Fig. 3a inset). Presence of continuous iron core is another unique feature of the MWNTs grown in the present experiments. Selected area diffraction patterns were obtained from various regions of the core. A typical diffraction pattern (along [012] zone axis) is shown in the inset of Fig. 5(a). At the highest magnification, the graphite spacings of 3.4 Å of the multiwalled carbon nanotube can be clearly resolved (Fig. 5(b)). TEM studies on other DQC substrates namely $Al_{67}$ $Co_{15}$ $Cu_8$ $Ni_{10}$ and $Al_{67}$ $Cu_8$ $Ni_{15}$ also showed multiwalled tube structure.

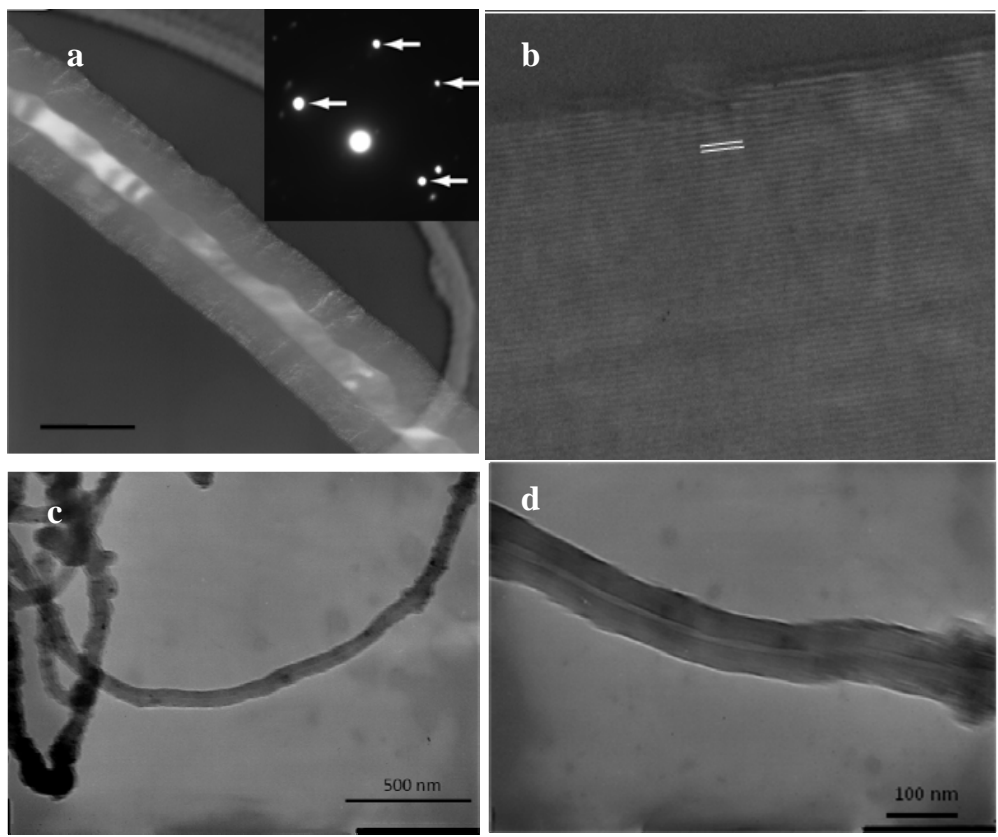

**Figure 5.** (a) TEM image of a MWNT grown on showing the presence of continuous iron filled core of the tube (scale = 50 nm). (Inset) A typical diffraction pattern (along [012] zone axis) from the iron filled area is presented showing its crystalline nature. (b) HRTEM showing the well graphitized walls of the MWNTs. The fringe of the graphite lattice shown is 3.334 Å. (c) TEM image of an individual MWNT on $Al_{67}$ $Co_{15}$ $Cu_8$ $Ni_{10}$ DQC substrate. The stunted growth and presence of catalyst impurities can be clearly seen. (d) High magnification TEM image of the same showing perfect multiwalled tube structure with hollow inner core and fringes between walls.

However, a large presence of catalysts in the vicinity was seen. Also the tubes were not well aligned and shorter in length as seen in Fig. 5(c). Fig.5 (d) shows a high magnification TEM image of a typical MWNT grown on $Al_{67}$ $Co_{15}$ $Cu_8$ $Ni_{10}$. It was observed that MWNTs on all DQC substrates have an average outer diameter of 60-70 nm and an average inner diameter of 10-16 nm. Thus the number of walls ranges approximately from 70-80. This diameter is larger than MWNTs grown on $SiO_2$. The catalyst encapsulation was however not observed in nanotubes grown on $SiO_2$ substrates. Fig. 6 shows the TEM images of MWNTs grown on $SiO_2$ substrates. The TEM images of nanotubes grown on $SiO_2$ show that the tubes are quite long and have an inner diameter of about 10-16 nm and an outer diameter of 42-47 nm. Assuming the interplanar distance between 2 graphitic planes as 0.334 nm the average number of walls in MWNTs synthesized on $SiO_2$ is 45-50 walls. Fig. 6 (a) shows a typical MWNT bundle grown on $SiO_2$. As mentioned above it was observed that the MWNTs are hollow and catalyst free to a large extent indicating good quality of growth. Fig. 6 (b) shows the structure of individual MWNTs at very high magnification. As mentioned earlier the deposit on IQC substrates was thought to be amorphous carbon which was indeed the case. This was unambiguously proved by HRTEM. Fig. 6(c) shows a HRTEM image of the deposit on IQC showing amorphous structure and diffraction pattern in Fig. 6(d) showing diffuse rings which are characteristic of amorphous structure.

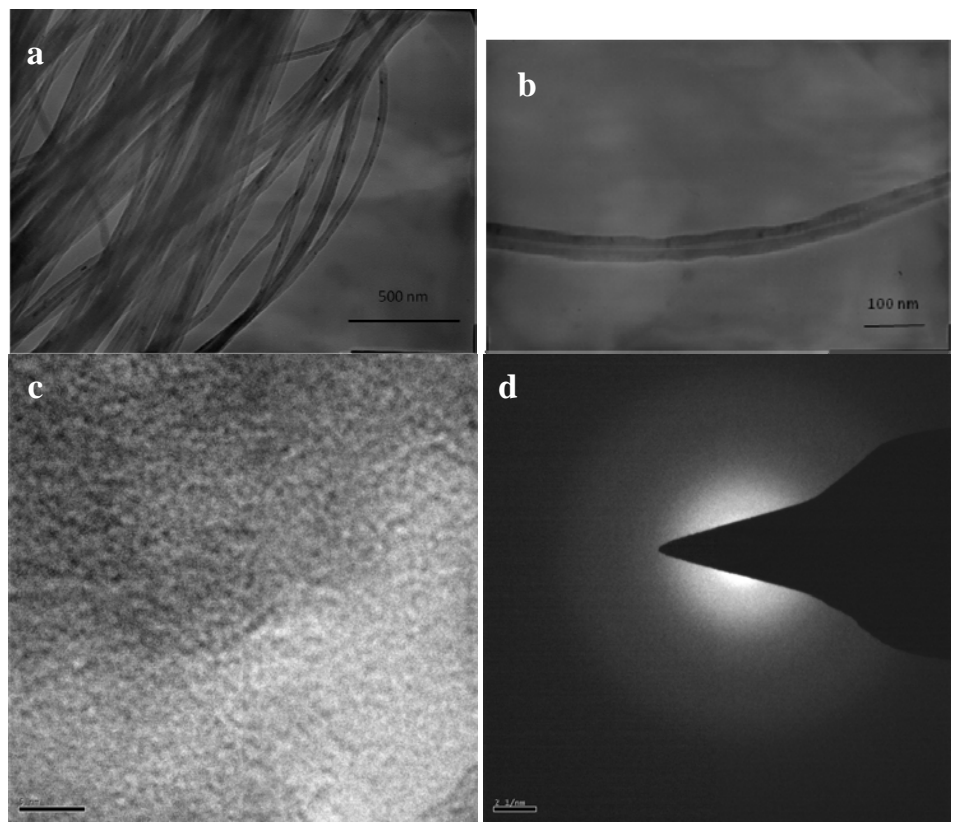

**Figure 6.** (a) TEM image of a MWNT bundle grown on Si/ 300nm SiO$_2$ substrate. The hollow, empty core and catalyst free tubes can be easily seen. (b) High magnification image of individual MWNT grown on Si/ 300 nm SiO$_2$ substrate. The inner hollow core can be easily seen in addition to the fringes due to multiwalled structure. (c)Bright field HRTEM image showing amorphous carbon deposition on the icosahedral quasicrystal. (scale bar = 5nm) (d) Diffraction pattern of the growth product on icosahedral quasicrystal. The pattern confirms that only amorphous carbon has been deposited and no growth of crystalline MWNTs took place.

We have also presented the magnetic measurement performed on the nanotubes grown on DQC rod. A squid magnetometer operating at room temperature was used for this measurement. The hysteresis loops obtained with the field applied parallel and perpendicular to the DQC rod is shown in Fig. 7. The saturation magnetic moment (M$_s$) for the samples was found to ~ 10$^{-4}$ emu and ~ 6X10$^{-4}$ emu for field applied parallel and perpendicular to the rod respectively. These values correspond to a total amount of ~ 10$^{-8}$ g iron present in the sample.

We have shown by the present study the possibility of growth of MWNTs on quasicrystalline substrate material. Experimental investigations have indicated that the mechanism for CNT growth on certain substrates using vapor phase catalyst delivery assisted process depends on the substrate-catalyst interaction. For example, the reason that growth occurs on SiO$_2$ and not on Si is due to a strong interaction of the catalyst with bare Si, leading to the formation of iron silicide and silicate[6]. One could generally assume that if the substrate shows a tendency to interact strongly with the metal catalyst particles, the CNT will not grow. Further, formation of high density catalyst nanoparticles (but not a continuous film) on the surface is an important criterion for growing vertically aligned carbon nanotubes on smooth substrates. This is only possible if there is minimal surface diffusion of the catalyst particles[14]. Therefore, we believe that in complex multicomponent intermetallic phases (quasi-crystals) due to the alloy constituents and their composition) the factors mentioned above balance out optimally to facilitate formation of stable catalyst particles resulting in aligned growth of CNTs, as opposed to other metals such as gold, copper etc. The aligned nanotube growth along with the simultaneous core filling of the nanotubes was a striking feature that was observed in our experiments. Though this behavior is not seen in nanotubes previously grown by CVD method on various other substrates, we believe that in the present situation the unusual catalytic behavior of the decagonal quasicrystal plays a crucial role and is responsible for the growth as well as the substantial core filling of the MWNTs with iron. On basis of this observation we also hypothesize and conclude that the quasicrystalline substrate-catalyst interaction should be weak since catalyst particles are frequently observed to be encapsulated within the nanotubes as compared to $SiO_2$-catalyst interaction. Therefore there is weak adhesion between the catalysts and quasicrystals thereby leading to encapsulation. The larger diameter of MWNTs on quasicrystalline substrates as compared to the ones grown on $SiO_2$ also indicates more iron diffusion on the quasicrystals which leads to larger catalyst size and consequently larger tube diameter. Better diffusion also indicates weak interaction of iron atoms/clusters with the quasicrystalline substrate.

## Conclusions

The present study showed the possibility of growth of MWNTs on decagonal quasicrystalline substrate material. The growth quality is comparable on both $SiO_2$ and quasicrystalline substrates except in the case of $Al_{68.85}Co_{15}Cu_{6.15}Ni_{10}/Al_{73.2}$ $Ni_{10.7}$ $Co_{16.1}$ and $Al_{67}$ $Co_{15}$ $Cu_8$ $Ni_{10}/$ $Al_{67}$ $Co_{15}$ $Cu_8$ $Ni_{15}$ where the inferior quality is mainly due to unwanted fluctuations in experimental conditions. No nanotube growth was observed on ternary icosahedral substrates. We believe this is due to the lower temperature of the substrate. Quaternary icosahedral phases, which are stable up to relatively high temperatures can exhibit amorphous carbon nanofiber/nanorod like growth with helical morphology.

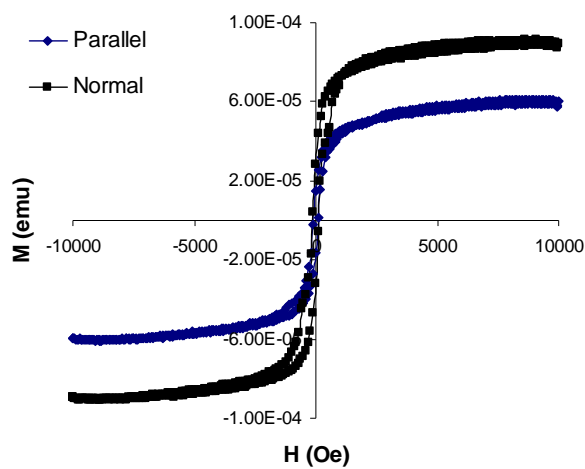

**Figure 7.** The hysteresis loops obtained with the field applied parallel and perpendicular to the DQC rod with nanotubes attached to it.

This presents with an interesting opportunity to synthesize novel 1D carbon nanomaterials for applications like hydrogen storage. Further investigations are needed with ternary icosahedral phases which are stable at and above the growth temperature quaternary quasicrystalline alloys without any iron, cobalt or nickel as constituent, are needed to see if the presence of these elements in the substrate material is indeed necessary for catalytic activity.

## Acknowledgements

One of the authors (GVSS) wishes to express his deep sense of gratitude to Infosys Foundation for the award of Dr. R. H. Kulkarni Memorial Visiting Fellowship as a result of which this study could be undertaken at RPI. He is thankful to the Dept. of Materials Engineering, RPI for providing the facilities. Helpful discussions with Prof. S.Lele gratefully acknowledged.The Al-Pd-Mn-Ga IQC quasicrystal was provided by Dr. Vincent Fournee and group at Ecole des Mines, Nancy, France for which he is gratefully acknowledged. The authors declare no competing interests.